\documentclass[prb,twocolumn,showpacs,preprintnumbers,amsmath,amssymb]{revtex4}

\usepackage{graphicx}
\usepackage{dcolumn}
\usepackage{bm}

\begin{document}

\title{Possible Superconductivity in Fe-Sb Based Materials:
Density Functional Study of LiFeSb}

\author{Lijun Zhang$^{1}$}
\author{Alaska Subedi$^{1,2}$}
\author{D.J. Singh$^{1}$}
\author{M.H. Du$^{1}$}

\affiliation{$^{1}$Materials Science and Technology Division,
Oak Ridge National Laboratory, Oak Ridge, Tennessee 37831-6114}
\affiliation{$^{2}$Department of Physics and Astronomy,
University of Tennessee, Knoxville, TN 37996}

\date{\today}

\begin{abstract}
We investigate the electronic and other properties of the hypothetical compound
LiFeSb in relation to superconducting LiFeAs and FeSe using density
functional calculations.
The results show that LiFeSb in the LiFeAs
structure would be dynamically stable in the sense
of having no unstable phonon modes, and would
have very similar electronic
and magnetic properties to the layered Fe based superconductors.
Importantly, a very similar structure for the Fermi surface and
a spin density wave related to but stronger than that in the corresponding
As compound is found.
These results are indicative of possible superconductivity analogous
to the Fe-As based compounds if the spin density wave can be suppressed
by doping or other means.
Prospects for synthesizing this material in pure form
or in solid solution with FeTe are discussed.
\end{abstract}


\maketitle

The finding of high temperature superconductivity  ($T_c\sim$26K) in
electron-doped LaFeAsO$_{1-x}$F$_{x}$,\cite{kamihara}
has resulted in widespread interest and exploration of related
materials, some of which have $T_c$ exceeding 55K.
In particular, superconductivity
has been found
in iron based oxy-arsenides by replacing La with other rare-earth metals,
\cite{wang-c,ren-epl,sefat,chen-prl,wen-epl} as well as oxygen-free arsenides
such as doped
BaFe$_2$As$_2$, \cite{rotter-prb,rotter} SrFe$_2$As$_2$,\cite{chen-cpl}
CaFe$_2$As$_2$,\cite{ni,torikachvili} and LiFeAs \cite{wu-g,pitcher,tapp}.
The common structural feature of this family of materials is the appearance
of Fe-As layers. These consist of an Fe square planar sheet tetrahedrally
coordinated by As atoms from above and below.
In addition, superconductivity occurs in doped LaFePO,
\cite{KamiharaY,liang,mcqueen}
although with a
lower $T_c$ and in PbO structure $\alpha$-FeSe$_{1-x}$.
\cite{hsu,yeh,mizuguchi}
These latter compounds also feature an Fe square lattice and a tetrahedral
coordination of the Fe, though not with As.
Importantly, the critical temperature of FeSe$_{1-x}$ increases strongly
with either Te substitution
\cite{yeh} or pressure, reaching 27 K. \cite{mizuguchi}
This high value of $T_c$ under pressure implies a relationship with
the Fe-As superconductors, which is also supported by
similarities of the properties and theoretical studies. \cite{subedi}
At present there is strong interest in finding new high temperature
Fe-based superconductors and especially in finding materials with
higher critical temperature.

One obvious direction is to examine antimonides. This is motivated by the
fact that the
properties of LaFePO and LaFeAsO appear to be closely related,
suggesting a similar mechanism of superconductivity, and furthermore
the compound with the heavier pnictogen (As) has the higher $T_c$ when doped.
However,
this is highly non-trivial from a chemical perspective because
Sb has a strong tendency to form Sb-Sb bonds in compounds.
This leads to a strong tendency for transition metal compounds of Sb to
contain more Sb than transition metal, as for example in skutterudite
CoSb$_3$ and LaFe$_4$Sb$_{12}$ or marcasite structure FeSb$_2$, although
FeSb is a known phase. \cite{kjekshus}
One way forward is provided by noting the structural similarity of LiFeAs with
PbO structure FeSe$_{1-x}$ and FeTe$_{1-x}$.
The chalcogenides, whose chemical formulas should more correctly be written
as Fe$_{1+x}$Se and Fe$_{1+x}$Te, occur in a tetragonal structure
with spacegroup $P4/nmm$ similar to LiFeAs, and consist of an Fe square lattice
tetrahedrally coordinated with Se/Te ions, the same as in the structure of the
Fe-As superconductors. \cite{chiba,finlayson,gronvold,leciejewicz}
These chalcogenides form with excess Fe, which occurs in a partially
filled $2c$ site, in particular the cation site
forming formally an enlarged tetrahedron 
around the Fe and approximately
five-fold coordinated by Te. \cite{finlayson,gronvold,leciejewicz}
This is the same site that is occupied by Li in LiFeAs.
Therefore, there is a close structural similarity between LiFeAs
and the chalcogenides Fe$_{1+x}$Se and Fe$_{1+x}$Te.
In particular the structure of LiFeAs is obtained by allowing full filling
of the $2c$ cation site with Li$^+$ and replacement of Te$^{2-}$ by As$^{3-}$.
Therefore we focus on hypothetical LiFeSb since it may be possible to
form it, or at the very least
some range of solid solution between Fe$_{1+x}$Te and
LiFeSb should be experimentally accessible, especially
considering that alloys of related phases containing Te and Sb typically
form as in e.g. the Bi-Sb-Te, AgSbTe-PbTe and AgSbTe-GeTe thermoelectrics,
and also that there are many known Zintl type phases based on Li, Sb and
metal atoms.

\begin{figure}
\includegraphics[width=3.4in]{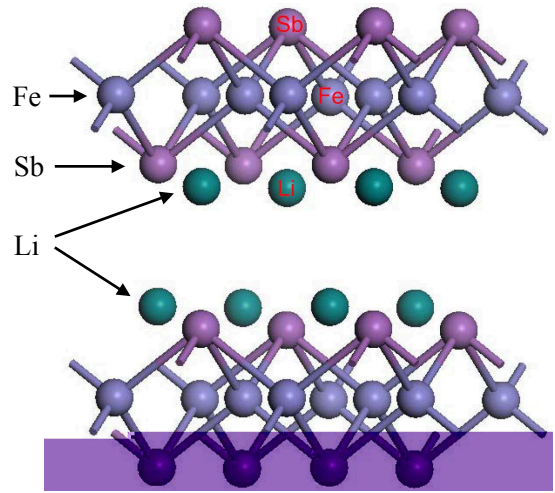}
\caption{(Color online) Crystal structure of hypothetical
LiFeSb with the relaxed structural parameters from
LAPW-LDA total energy minimization. Note that while for clarity similar
size spheres are used for the different atoms, from a crystal
chemical point of view Sb$^{3-}$ anions are very large while
Li$^+$ is very small.}
\label{stru}
\end{figure}

The crystal structure of LiFeSb is assumed to be isostructural with
LiFeAs with the space group of $P4/nmm$.\cite{wu-g,pitcher,tapp}
As shown in Fig. \ref{stru}, the Fe-Sb layers formed by edge-shared
tetrahedral FeSb$_4$ units are alternately spaced along the $c$-axis direction,
and intercalated with Li.
Note that the Li, which occurs as Li$^{+}$, is coordinated
by Sb.
The structural parameters were calculated by local density
approximation (LDA) total energy minimization with the full-potential
linearized augmented plane wave (LAPW) method. \cite{singh2006}
The calculated tetragonal lattice parameters are
$a$ = 4.0351 \AA, $c$ = 6.3712 \AA,
and internal coordinates
Li(2$c$) (0.25,0.25,0.697), Fe(2$a$) (0.75,0.25,0), Sb(2$c$) (0.25,0.25,0.228).
The Fe-Sb bond length is 2.486 \AA,
slightly larger than 2.4204 \AA\ for LiFeAs,\cite{tapp}
which might be attributed to the larger size of the Sb$^{3-}$ anion relative
to As$^{3-}$.
The Fe-Fe distance is 2.853\AA,
also a bit larger than the corresponding value (2.6809\AA) in LiFeAs,
but still short enough for direct Fe-Fe interaction.

The electronic structure and magnetic properties calculations were performed
within LDA-LAPW method,
similar to previous reports.\cite{singh-du,singh,subedi}
LAPW sphere radii of 1.8 $a_0$, 2.0 $a_0$, and 2.1 $a_0$ were used for
Li, Fe and Sb, respectively.
Converged basis sets were used. These consisted
of LAPW functions with a planewave
cutoff determined as $R_{\rm Li}k_{max}$=8.0 plus local orbitals
both to relax linearization and to include the semicore states.
The zone sampling for the self-consistent calculations was done
using the special {\bf k}-points method, with a 16x16x8 grid. Finer
grids were used for the density of states and Fermi surface.
The lattice dynamical properties were calculated through the frozen
phonon method \cite{togo} (or small displacement method\cite{alfe}).
The required forces were obtained through the projector augmented-wave
(PAW) method \cite{kresse} in VASP code,
within the generalized gradient approximation of
Perdew, Burke and Ernzerhof (PBE).\cite{pbe}
We also fully relaxed crystal structure and
calculated electronic structures with PBE-PAW method and the results show
excellent agreement with those by LDA-LAPW method
(with a remarkably small maximum discrepancy of 0.3\% in structural parameters).
This cross-checking supports the reliability of the
calculations and consistency of the different methods employed.
We emphasize that these calculations for the structure were done without
the inclusion of magnetism.

As discussed in detail in Ref. \onlinecite{mazin-mag},
there is generally a substantial underestimation of the pnictogen heights
in these compounds when
calculations are done in this way, while on the other hand
in magnetic calculations LDA and GGA results differ, with the GGA
giving much larger magnetic moments than experiment as well as
magnetism that persists throughout the phase diagram in disagreement
with experiment.
LDA calculations done with the GGA structure give an intermediate
state less magnetic than the GGA calculations but more magnetic
than experiment, while in magnetic LDA calculations the As height is
still substantially underestimated.
In fact, LDA calculations done at the non-magnetic LDA As height (which
agrees with the GGA As height) give the weakest magnetism, closest to
experiment, though still overestimating the strength of the SDW.
It was conjectured that these problems are
consequences of strong spin fluctuations.

This conjecture is supported both by comparison of theoretical
results with experiment as well as experimental observations, such as
the highly unusual
increasing with $T$ susceptibility,
$\chi(T)$ above the spin density wave
ordering temperature observed in some compounds.
\cite{mcguire,klauss,klingeler}
Such an increasing shape
suggests that strong magnetic correlations with
the character of the SDW persist well above $T_N$; since this itinerant
magnetic state is driven
by electrons at the Fermi energy a related electronic reconstruction of the
Fermi surface may also be expected above $T_N$; this may be seen
perhaps in photoemission.
This shape of $\chi(T)$ persists also in the normal state for
doped samples, where there is superconductivity but no SDW. \cite{klingeler}
Other evidence for strong spin fluctuations
comes from core level spectroscopy
\cite{bondino} and transport data showing strong scattering
above the ordering temperature. \cite{mcguire}
Returning to the increasing $\chi(T)$, which continues up to high
temperature, one interesting possibility is that this represents
preformed pairs that can condense into either superconducting or 
SDW order and which begin forming at a very high non-observed temperature
above which $\chi(T)$ would return to a more normal decreasing with $T$ shape.
In any case, in this work where we compare the compounds, we consistently used
the relaxed atomic coordinates from non-magnetic calculations.

\begin{figure}
\includegraphics[width=3.7in]{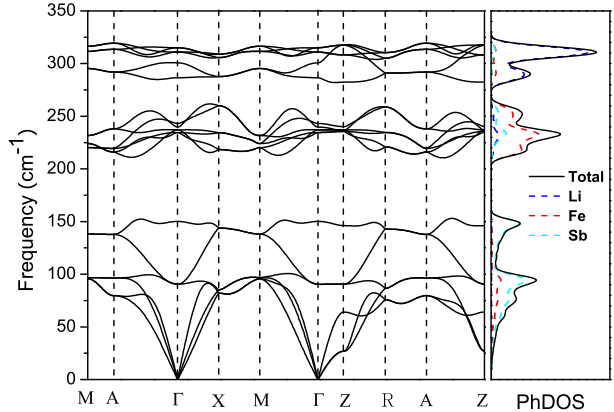}
\caption{Left panel: Calculated phonon dispersion curves for LiFeSb.
Right panel: (Color online) The total and projected (onto atoms) phonon DOS.}
\label{phon}
\end{figure}

A requirement for a compound to be made is that the lattice be stable.
We verified that this is the case for hypothetical LiFeSb by
calculating the vibrational modes of the compound.
\cite{baroni}
We find no soft or unstable modes and no soft elastic constants.
The calculated phonon dispersion curve and phonon DOS for LiFeSb are shown in
Fig. \ref{phon}.
Due to larger difference in atomic weights compared to LiFeAs
the phonon spectrum of LiFeSb is divided into three separated manifolds.
The region of high frequencies (above 275 cm$^{-1}$)
is dominated by Li,
while the moderate (between 200 and 275 cm$^{-1}$) and
low (below 150 cm$^{-1}$) frequency manifolds
mainly derive from Fe and Sb respectively.
As may be seen, all the phonon frequencies are safely positive and there are
no optical phonon branches with dispersions that dip towards zero frequency.
This shows
that the $P4/nmm$ structure of LiFeSb is dynamically stable.
We also calculated the heat of formation from the elements. We obtain
-0.51 eV per formula unit (i.e. -49 kJ/mole formula unit),
which indicates that the compound may be delicate
but would at least be stable against
decomposition into elements.
Therefore we continue to discuss the magnetic and electronic properties.

\begin{figure}
\includegraphics[width=2.5in,angle=270]{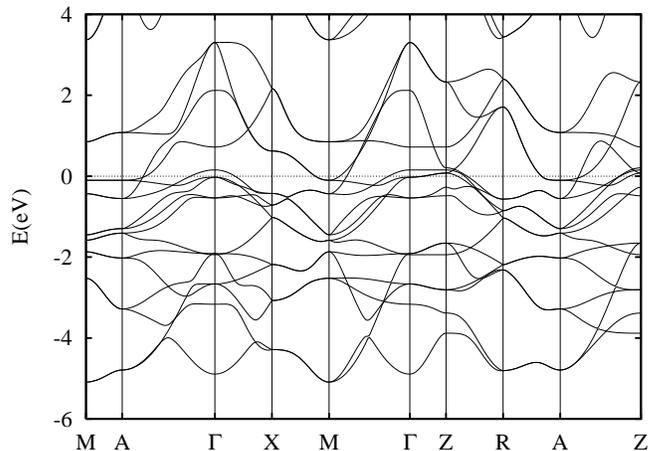}
\caption{Calculated LDA band structure of LiFeSb using the calculated
structural parameters. The Fermi energy is at 0 eV.}
\label{band}
\end{figure}

\begin{figure}
\includegraphics[width=2.5in,angle=270]{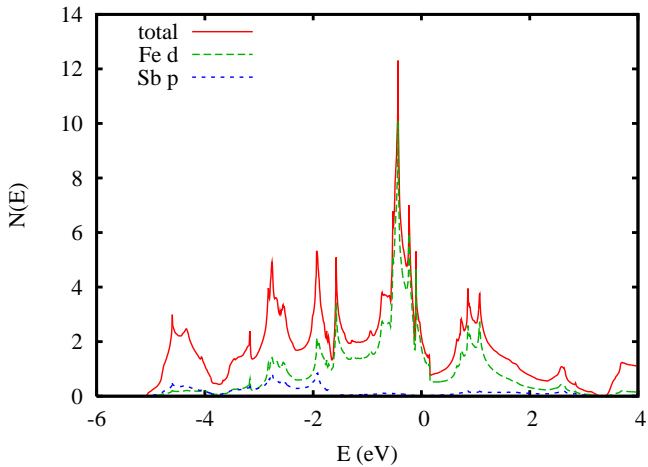}
\caption{(Color online) Calculated total and partial electronic DOS for LiFeSb, on a per formula unit basis. The contribution from Li-2$s$ state lying in deep energy range was not shown. The projections are onto the LAPW spheres, thus the Sb-5$p$ was slightly underestimated owing to its more extended orbitals.}
\label{dos}
\end{figure}

\begin{figure}
\includegraphics[width=3.5in]{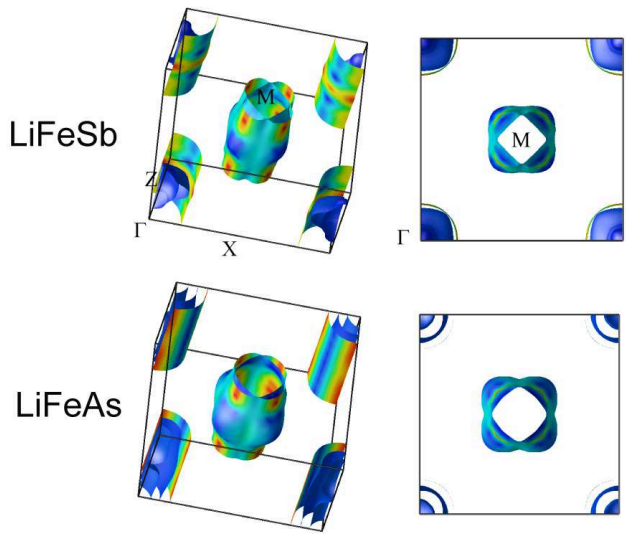}
\caption{(Color online) Calculated LDA Fermi surface of LiFeSb
in comparison with LiFeAs,
shaded by band velocity with blue as low velocity.
The right panels are top views along the $c$-axis direction.}
\label{fermi}
\end{figure}

Our main results for the electronic structure of LiFeSb,
are given in Figs. \ref{band}, \ref{dos} and \ref{fermi},
which show the calculated band structure, electronic density of states (DOS),
and Fermi surface, respectively.
The general shape of band structure
near the Fermi energy $E_F$ is very similar to the calculated results
for LiFeAs.\cite{nekrasov-lifeas,singh}
There are compensating heavy hole and electron Fermi surfaces,
with two electron cylinders at the zone corner ($M$) and hole
surfaces around the zone center.
The hole surfaces consist of 2D cylindrical and small heavy 3D sections.
Similar to the Fe-As based materials,
\cite{singh-du,nekrasov,ma-f,ma2,nekrasov-lifeas,singh}
the electron Fermi surface of LiFeSb may be described as two intersecting
cylindrical sections of elliptical cross-section, with major axes at 90$^\circ$
to each other and centered at the $M$ point.
We find somewhat a different hole Fermi surface structure from LiFeAs,
with only one complete hole cylinder at the zone center,
along with two additional heavier 3D hole pockets.
It can be seen that electron cylinders are more two dimensional
than in LiFeAs.
Also, the 2D hole cylinder is close in size to that of the
electron cylinders, which may be expected to lead to nesting.
Thus there is strong nesting of Fermi surface at the two-dimensional (2D)
nesting vector ($\pi$,$\pi$).
This would be expected to lead to an SDW state related to the $M$ point,
as in the Fe-As based superconductors.
\cite{dong-j,mazin,ma-f,ma2,yildirim,yin-prl}
We studied the energetic stability of SDW state for LiFeSb
directly using a doubled cell containing
lines of Fe atoms with parallel spin in the Fe-Sb layers and do in fact
find a stable SDW state. Within the LDA with the LDA structural parameters,
the local spin moment of the SDW state is 1.12 $\mu_B$,
much larger than the corresponding value (0.69 $\mu_B$) for
LiFeAs calculated in the same way,\cite{singh}
indicating that LiFeSb has a more stable SDW.

The qualitative similarity to the electronic structure
of the Fe-As based superconductors
\cite{singh-du,nekrasov,ma-f,ma2,nekrasov-lifeas,singh}
is also evident in the DOS.
The Sb $p$ states are located mainly below $-1.7$ eV relative to
$E_F$, and are only
moderately hybridized with the Fe $d$ states, indicating that the Sb
is anionic with valence close to 3.
The DOS near the Fermi level is dominated by Fe $d$ states,
deriving from the metallic Fe$^{2+}$ sublattice with direct Fe-Fe interactions,
and has a characteristic pseudogap near $E_F$.
In fact, $E_F$ lies on the low energy side of the pseudogap,
where $N(E_F)$ is decreasing with energy but still high.
Specifically, the value of $N(E_F)$ = 2.2 eV$^{-1}$ per Fe both spins
is much larger than that for LiFeAs and is comparable to the
oxy-arsenides (e.g. $N(E_F)$ calculated in the
same way for LaFeAsO is 2.6 eV$^{-1}$),
which are the Fe-As compounds with higher $T_c$.
For comparison, the values for LiFeAs and BaFe$_2$As$_2$ are
1.79 eV$^{-1}$ and 1.53 eV$^{-1}$ on a per Fe basis, repectively, when
calculated in the same way.\cite{singh}

Within the Stoner theory,
the appearance of an instability of the paramagnetic state towards
itinerant ferromagnetism would be determined by the criterion $N(E_F)I > 1$,
where $I$ is the Stoner parameter, with the typical value in Fe compounds of
$I \sim$ 0.7 -- 0.8 eV.
Thus, the significantly higher $N(E_F)$ in LiFeSb would inevitably
place it closer to magnetism in general than LiFeAs or BaFe$_2$As$_2$.
While the mechanism of superconductivity has yet to be established,
there is accumulating evidence of a connection with magnetism,
and so chemically tuning the proximity to magnetism is a likely strategy for
modifying the superconductivity.
In general, the Fe-based superconductors
exhibit temperature-induced magnetic and structural
phase transitions with a spin density wave (SDW) character related to the
Fermi surface nesting.
\cite{delacruz,nomura,rotter-prb}
Superconductivity appears as the spin density wave is suppressed
by doping or pressure.

Electronic structure calculations
\cite{singh-du,nekrasov,ma-f,ma2,nekrasov-lifeas,singh}
show that all these materials have compensating small
electron and hole Fermi
surfaces, with nesting between 2D electron sheets and heavier 2D
hole sheets, which are separated by ($\pi$,$\pi$).
This is associated with the SDW magnetic state.
\cite{dong-j,mazin,ma-f,ma2,yildirim,yin-prl}

Within this framework, the idea that going to heavier ligands may
be beneficial for superconductivity is supported by previous
density functional calculations. These have shown that the electronic
structures of the Fe-As superconductors are rather ionic with the
exception of the Fe layers, which are metallic due mainly to Fe-Fe
interactions. \cite{singh-du}
This is different from the cuprates where hopping is through the
O atoms in the CuO$_2$ planes, and implies that the ligand (O/As) atoms
play a less crucial role in the properties of the FeAs superconductors
than in the cuprates.
Furthermore, it has been found that there is a strong connection
between the As position above the Fe plane and the magnetic properties,
with higher positions yielding stronger magnetism. \cite{yin-prl,mazin-mag}
This is supported by calculations comparing FeSe and FeTe \cite{subedi}
and for hypothetical LaFeSbO in comparison with LaFeAsO. \cite{moon}
In both cases stronger magnetism is found in going to the larger ligand,
which because of its size is then further from the Fe plane yielding
narrower bands and higher $N(E_F)$.
The combination of a stronger SDW and higher $N(E_F)$ leading to stronger
spin fluctuations in general and in particular away from the nesting
vector may be crucial. This is because the ordered SDW is antagonistic
to superconductivity. In scenarios where the associated spin-fluctuations
that couple the electron and hole Fermi surface sections play the main role
in pairing, as discussed in Refs. \onlinecite{mazin} and \onlinecite{kuroki},
spin fluctuations away from the nesting vector while not directly pairing
may play a very important role. This is because they would compete with the SDW
preventing long range order and leading to a renormalized paramagnetic state
even though in mean field the SDW may be the predicted ground state as
in the LDA. In any case, in these scenarios the role of doping is to
weaken and broaden the peak in the susceptibility associated with the nesting,
destroying the SDW in favor of a state with spin fluctuations around the zone
corner. We also note that in case the SDW is not destroyed by doping alone,
it may be possible for it to be destroyed by disorder yielding
superconductivity in an alloy system such
as Fe$_{1+x}$Te -- LiFeSb. This may be possible because
the SDW is related to
a divergence in the peak of the susceptibility, $\chi({\bf q})$, while
within a spin-fluctuation mediated framework
superconductivity will in general be related not to the peak value but
to an integral over the Fermi surface, i.e. a high average
value over some region of the zone, specifically the region for which
the wavevector connects the electron and hole Fermi surfaces.
\cite{mazin,fay}
Also near divergences
will be pair breaking for superconductivity.
This may also partly explain why superconductivity in these
phases is relatively robust against alloying with Zn or Co in the Fe
planes,
\cite{sefat-co1,sefat-co2,li-zn}
even though in an unconventional superconductor
scattering, including non-magnetic scattering, is pair breaking.
Thus, even if the magnetic ground state cannot be destroyed by doping in
Fe$_{1+x}$Te or LiFeSb (supposing that this can be synthesized) it may
be destroyed in favor of superconductivity in the solid solution between
these two compounds.

In any case, our results show that, if it can be synthesized,
LiFeSb will have electronic and magnetic properties closely related to those
of the Fe-As based superconductors, and in particular will show a qualitatively
similar Fermi surface structure and tendency towards an SDW state.
In comparison with LiFeAs, it will have a higher $N(E_F)$ and a stronger SDW.
This may favor higher critical temperatures, at least within a scenario
with interband pairing mediated by spin fluctuations associated with the 
Fermi surface nesting.
In addition, this material is found to be dynamically stable,
evidenced by the absence of any unstable phonon modes.
It is also worth noting that this solid solution contains no elements
as toxic as As.
As such it would be of considerable interest to attempt synthesis of this
compound or its solid solution with Fe$_{1+x}$Se, LiFeAs
and especially Fe$_{1+x}$Te.

\acknowledgements
We are grateful for helpful discussions with D. Mandrus, I.I. Mazin and
B.C. Sales.
This work was supported by the Department of Energy,
Division of Materials Sciences and Engineering.

\bibliography{LiFeSb}

\end{document}